\documentclass[12pt]{article}
\usepackage{graphicx}
\usepackage{amssymb}
\usepackage{epstopdf}
\usepackage{multirow}
\newcommand\mathC{\mkern1mu\raise2.2pt\hbox{$\scriptscriptstyle|$}
        {\mkern-7mu\rm C}}              %%% The complex  numbers
\newcommand{\mathR}{{\rm I\! R}}         % The real numbers

\newcommand{\be}{\begin{equation}}
\newcommand{\ee}{\end{equation}}

\let\ssection=\section
\renewcommand{\section}{\setcounter{equation}{0}\ssection}

\newcommand\bi{\begin{itemize}}
\newcommand\ei{\end{itemize}}

\newcommand{\al}{\ensuremath{{\alpha}}}

\begin{document}
\begin{center}
{\large\bf On Time in Quantum Physics}
\end{center}

\begin{center}
Jeremy Butterfield\\
Trinity College, Cambridge CB2 1TQ: jb56@cam.ac.uk
\end{center}

\begin{center}
Published in {\em The Blackwell Companion to the Philosophy of Time}, edited by A. Bardon and %%@
H. Dyke, Wiley-Blackwell (2013), pp. 220-241.
\end{center}


\begin{center} Tues 19 June 2012 \end{center}

\begin{abstract}
First, I briefly review the different conceptions of time held by three rival %%@
interpretations of quantum theory: the collapse of the wave-packet, the pilot-wave %%@
interpretation, and the Everett interpretation (Section 2).

Then I turn to a much less controversial task: to expound the recent understanding of the %%@
time-energy uncertainty principle, and indeed of uncertainty principles in general, that %%@
has been established by such authors as Busch, Hilgevoord and Uffink.

Although this may at first seem a narrow topic, I point out connections to other conceptual %%@
topics about time in quantum theory: for example, the question under what circumstances %%@
there is a time operator (Section \ref{opors}).

\end{abstract}

\newpage
\tableofcontents

\newpage

\section{Introduction}\label{prosp}
Time, along with such concepts as space and matter, is bound to be a central concept of any %%@
physical theory,
especially of a basic or fundamental one. So how time is treated in a physical theory is  %%@
bound to be of philosophical interest: and all the more so, if the theory in question is %%@
our current best theory and is philosophically interesting or even problematic---as quantum %%@
theory evidently is. Besides, when one considers the philosophical problems about quantum %%@
theory---especially non-locality and the measurement problem---and how the various %%@
interpretations of the theory propose to solve them, one soon sees that the interpretations %%@
have distinctive consequences for the nature of time. The details are in Section %%@
\ref{noroads}.

But this situation suggests a deflating thought. Namely: that a conclusive discussion of %%@
time in quantum theory would require a resolution of the debate between those %%@
interpretations. And furthermore, in the absence such a resolution, the discussion is %%@
liable to turn into a survey of the {\em pros} and {\em cons} of the various %%@
contenders---in which the discussion about time would be likely to occur, not `up front', %%@
but briefly at the end, in an enumeration of the interpretations' distinctive consequences %%@
about time.

This tendency, for reflection to turn towards the general topic of interpreting quantum %%@
theory, is aggravated by another deflating line of thought. Namely: when we set aside %%@
non-locality and the measurement problem, and ask how quantum theory's {\em formalism} %%@
treats time, the answer is, broadly speaking: quantum theory has nothing very special to %%@
say about time---it just tags along with whatever is said by the cousin classical theory, %%@
be it Newtonian or relativistic. Again, I will spell out this line of thought in Section %%@
\ref{noroads}. For now, the point is just that this line of thought also suggests that %%@
`time in quantum physics' is, in the present state of knowledge, not much of a %%@
topic---except as a corollary to whatever one concludes about various large and unresolved %%@
issues about the dreaded quantum.

So what to do? Happily, there are several topics about time in quantum physics, that are %%@
largely independent of the issues which these two deflating lines of thought focus on---and %%@
that are significant in their own right. I will pursue one such topic, viz. the time-energy %%@
uncertainty principle, emphasizing how it relates to general results about uncertainty %%@
principles. At first sight, this is bound to seem a very narrow topic. But in fact, it is %%@
not; for the following two reasons.\\
\indent (1): Of course, {\em uncertainty principles} lie at the heart of quantum physics. %%@
But it turns out that the textbook tradition in quantum theory gives a limited, indeed %%@
downright misleading, view of them: defects which have especially blighted the %%@
understanding of the time-energy uncertainty principle. Fortunately, the situation has been %%@
much clarified in the last twenty-five years. So it is worthwhile to explain the recent %%@
progress from a philosophical viewpoint: which I propose to do from Section \ref{UPgenl} %%@
onwards. I shall concentrate on just a trio of authors (Busch,  Hilgevoord and Uffink): I %%@
simply refer the reader to the references in their papers, in order to properly appreciate %%@
the wider community of research and scholarship about the various issues.\\
\indent (2): As we shall see, uncertainty principles shed light on other topics about time. %%@
For example, one such topic is {\em time operators}: since in quantum theory, physical %%@
quantities such
as position are represented by operators, the question arises whether one can (and should!) %%@
define a time operator---and if not, why not. The discussion of this goes back to the %%@
founding fathers of quantum theory, especially Pauli: a discussion which, like that of %%@
uncertainty principles in general, has been beset by confusion and misunderstanding. I will %%@
not go into this in detail, but will touch on it in Section \ref{3times}.\footnote{So much %%@
by way of justifying my focus on uncertainty principles. For a fine recent anthology which %%@
covers this and many other issues about time in quantum physics, cf. Muga et al. (2008). As %%@
an example of progress since then, I recommend recent analyses of times of arrival, sojourn etc. of a quantum system in a spatial region. Thus for analyses in the decoherent histories %%@
framework, cf. Halliwell (2011), Yearsley (2011). For a recent time-energy uncertainty principle for an arrival time, cf. Kiukas et al. (2011).}

The plan of the paper is as follows. In Section \ref{noroads}, I briefly sketch some %%@
options about the interpretative issues I will {\em not} pursue. Then in Section %%@
\ref{UPgenl}, I survey my three chosen authors' clarifications of uncertainty principles in %%@
general. Then I follow Busch (1990 Section 2, 2008 Section 2) in distinguishing three roles %%@
for time in
quantum physics (Section \ref{3times}). The main role will be called `intrinsic time' %%@
(Section \ref{intrinsic}); but I will also briefly discuss the others, in particular time %%@
operators (Section \ref{opors}). Then in the closing Section \ref{UPintrinsic}, I discuss %%@
time-energy uncertainty principles with intrinsic times.

\section{Some roads not taken}\label{noroads}
I will spell out the two deflating lines of thought with which Section \ref{prosp} began %%@
(Sections \ref{timeinterp} and \ref{nodifference}). Then I will give a few
references about another conceptual issue I  neglect, viz. time-reversal (Section %%@
\ref{reverse}).

\subsection{Time in the interpretations of quantum theory}\label{timeinterp}
Notoriously, the interpretation of quantum theory is contested. There are two main %%@
problems: quantum non-locality and the quantum measurement problem, of which the second is %%@
generally regarded as more fundamental. This is not the place to expound, or even sketch, %%@
these interpretations, since assessing them requires considerations, both technical and %%@
conceptual, that are largely independent of time. But fortunately, we can easily see, %%@
without going into any details, that the resolution of the debate between them will have %%@
major repercussions for our conception of time.

I will confine myself to the following three well-known interpretations, or rather families %%@
of more precise interpretations: the collapse of the wave-packet, the pilot-wave %%@
interpretation, and the Everett or many-worlds interpretation.  I will describe how each %%@
has distinctive consequences about the nature of time: and I will discuss them in what I %%@
take to be the order of increasing radicalism of the consequences.

Two admissions at the outset. (1): Each of these families is a broad church. The `collapse %%@
of the wave-packet' will cover, not just a textbook-like minimal instrumentalism, but also %%@
the Copenhagen interpretation (whatever exactly that is!) and the dynamical reduction %%@
programme of Ghirardi, Pearle, Penrose and others. The `pilot-wave' will cover, not just de %%@
Broglie's and Bohm's original theories with point-particles always having a definite %%@
position, but also analogous theories, including field theories. And `Everett' will cover %%@
various proposals for how to define `worlds' or `branches'. But I will suggest that, %%@
despite the variety within each `church', its members will have broadly similar %%@
consequences about the nature of time. (2): I agree that `interpretation' is a contentious %%@
word. It is perhaps better to say that dynamical reduction and pilot-wave theories are %%@
rivals, or at least supplements, to quantum theory, rather than interpretations of it. But %%@
it will be harmless to keep to the jargon of `interpretation'.

\subsubsection{The collapse of the wave-packet}\label{collapse}
`The collapse of the wave-packet' refers to an irreducibly indeterministic change in the %%@
state of an isolated quantum system, contravening the deterministic and continuous %%@
evolution prescribed by the quantum theory's fundamental equation of motion (the %%@
Schroedinger equation). The collapse is postulated in order to solve the measurement %%@
problem: viz. by securing that measurements have definite outcomes, albeit ones that are %%@
not determined by the previous quantum state. Anyone who advocates such a collapse faces %%@
several questions. Three of the most pressing ones are as follows. Under exactly what %%@
conditions does the collapse occur?  What determines the physical quantity (in the %%@
formalism: the basis) with respect to which it occurs? How can the collapse mesh with %%@
relativity?

\indent These are of course questions for the foundations of physics, not philosophy. But %%@
however they are answered, the proposed irreducible indeterminism raises for philosophy the  %%@
question how exactly to understand the various alternative futures. In particular: is %%@
indeterminism compatible with the idea of a single actual future, i.e. the `block universe' %%@
or `B-theory of time'? Some say Yes (e.g. Lewis (1986, p. 207-208; Earman 2008, p. 138 %%@
footnote 8). Broadly speaking, the idea is:  an indeterministic event requires that two or %%@
more possible histories (possible worlds) match utterly up to the event in question, but %%@
thereafter fail to match (called: `diverge'), reflecting the differences in the event's %%@
various possible outcomes and in these outcomes' later causal consequences. Others say No %%@
(e.g. McCall 2000). The idea is:  an indeterministic event requires that a single possible %%@
history (possible world) splits at the event in question, with future branches %%@
incorporating the different outcomes and their respective causal consequences.

\indent Thus our topic, time in quantum physics,  runs up against notoriously hard %%@
questions in the metaphysics of modality.\footnote{It also runs up against the cluster of %%@
issues, dubbed `the direction/arrow of time'. But I set this aside: cf. Wallace (this %%@
volume).} How exactly should we conceive of possible histories? We presumably need to %%@
answer this in order to settle how we should understand their matching: as a matter of some %%@
kind of isomorphism or counterparthood, or as fully-fledged identity? Here, with relief, I %%@
duck out of these questions. But for recent work that relates them to physics and %%@
especially relativity, I recommend Earman (2008a): his Section 3 discusses how a wide %%@
variety of theorems about spacetime structures make it very hard  to make the branching %%@
view mesh with relativity; and his Section 2 discusses how the branching view relates to %%@
the work of the `Pittsburgh-Krakow' school of Belnap, Placek and others, on what they call %%@
`branching spacetimes' (developed with an eye on quantum theory, especially quantum %%@
non-locality, e.g. Placek (2000)). Cf. also Earman (2008) for an assessment, again in the %%@
light of modern physics, of the proposal (Broad 1923, Tooley 1997) that over time, reality %%@
`grows by the accretion of facts'.

\subsubsection{The pilot-wave}\label{pilot}
The pilot-wave interpretation adds to quantum theory's deterministic evolution of the %%@
orthodox quantum state (wave-function), the postulate that certain preferred quantities %%@
have at all times a definite value: which also evolves deterministically in a manner %%@
governed by the quantum state/wave-function.

\indent The original and best-studied version postulates that in non-relativistic quantum %%@
theory, the preferred quantity is the position of point-particles. These positions evolve %%@
deterministically according to a guidance equation that requires the particle's momentum at %%@
any time to be proportional to the gradient of the phase of the wave-function. (This %%@
equation is part of the orthodox formalism, albeit differently interpreted.) In other %%@
versions, other quantities, such as a field quantity like the magnetic field, are %%@
preferred, i.e. postulated to have definite values that evolve deterministically `guided' %%@
by the quantum state.

\indent But as the phrase `at any time' hints, most current versions of the pilot-wave %%@
interpretation use an absolute time structure (frame-independent simultaneity), as in %%@
Newtonian physics. Agreed, some of these versions, especially in field theory, secure the %%@
Lorentz symmetry characteristic of special relativity. But this is an approximate and %%@
emergent symmetry, governing a certain regime or sector of the theory, not a fundamental %%@
one.  (In a somewhat similar way, the pilot-wave interpretation recovers the apparent %%@
indeterminism of orthodox quantum theory (``the collapse of the wave packet'') as  an %%@
emergent feature due to averaging over the unknown, but deterministically evolving, %%@
definite values.)\footnote{For details, cf. Bohm and Hiley (1992, Chapter 12), Holland %%@
(1993, Chapter 12), Butterfield (2007, Appendix) and Struyve (2010, 2011). In fairness, I %%@
should add that most versions of Section \ref{collapse}'s dynamical reduction programme %%@
also use an absolute time, and it is hard to build a relativistic version: cf. e.g. %%@
Bedingham et al. (2011).}

\indent So if we follow the lead of these current versions, the consequences for the nature of time %%@
are striking: and  they are crucial for physics, not ``just'' philosophy. Namely: these %%@
versions resurrect the absolute time structure of Newtonian physics. For the philosophy of %%@
time, that would put the cat among the pigeons. After a century of accommodating our %%@
metaphysics and epistemology of time, one way or another, to relativity's denial of %%@
absolute time, we philosophers would be invited back to square one.

\subsubsection{Everett}\label{Everett}
The Everett or many-worlds interpretation proposes to reconcile quantum theory's %%@
deterministic evolution of the orthodox quantum state with the collapse of the wave packet, %%@
i.e. measurements having definite outcomes with various frequencies, by saying that %%@
measurement processes involve a splitting of the universe into branches. Obviously, this %%@
returns us to the murky issues, both physical and philosophical, glimpsed in Section %%@
\ref{collapse}: about the conditions under which a branching occurs, how branching can mesh %%@
with relativity, and how we should understand branching. So it is hardly surprising that %%@
this interpretation has traditionally been regarded as vaguer and more controversial than %%@
others. Thus Bell, in his masterly (1986) introduction to interpreting quantum theory, %%@
wrote that it `is surely the most bizarre of all [quantum theory's possible %%@
interpretations]' and seems `an extravagant, and above all extravagantly vague, hypothesis. %%@
I could almost dismiss it as silly' (pp. 192, 194).

\indent But since Bell wrote, Everettians have made major improvements to their %%@
interpretation. (Sad to say: Bell died in 1990, so that he did not engage with these %%@
developments.) In my opinion, the two main improvements have been: \\
\indent \indent (i): To combine the physics of decoherence with the philosophical %%@
(``functionalist'') idea that objects  in `higher-level' ontology, e.g. a cat, are not some %%@
kind of aggregate (e.g. a mereological fusion) of lower-level objects, but rather %%@
dynamically stable patterns of them. This suggests that  the proverbial Schroedinger's cat %%@
measurement involves an approximate and emergent splitting, after which there really are %%@
two cats (or two broad kinds of cat), since the total wave-function is peaked over two  %%@
distinctive patterns in the classical configuration space. More precisely, it is peaked %%@
over two kinds of pattern: the legs, tail and body all horizontal, still and cool (`dead'), %%@
and  the legs and tail vertical, moving and warm (`alive').  \\
\indent \indent (ii): To develop various arguments justifying, from an Everettian %%@
perspective, the orthodox (Born-rule) form of quantum probabilities.

Both these improvements are developed in detail by many papers over the last twenty years. %%@
Recent work includes the papers in Saunders et al. (2010), and Wallace (2012; 2012a, Sections 3, 4); (and %%@
the former includes critical assessments by non-Everettians). And both lead to fascinating %%@
open questions. But for our topic of time in quantum physics, it is (i) not (ii) that is %%@
relevant.

\indent Here the important point is how hard it is to get one's mind around the central %%@
idea. That is: the dizzying vision whereby decoherence processes yield a continual, but %%@
approximate and emergent, splitting of the universe that \\
\indent \indent (a) meshes fundamentally with relativity, so that there is no absolute time %%@
structure; and \\
\indent \indent (b) is to be combined with almost all objects---not just macroscopic %%@
objects like cats, tables and stars, but anything that classical physics successfully %%@
describes as having a spatial trajectory etc.: for example, large molecules---being treated %%@
as patterns. (Or better: being treated as the quantum state being peaked above such %%@
patterns in an abstract classical configuration space.)

In saying that the important point is how hard and dizzying is the vision, I do not mean to %%@
condemn it as extravagant or silly. I mean just that it is hard to think about---but that %%@
by no means makes it less alluring! Indeed, Everettians admit the difficulty, as well as %%@
the attraction. And some think there is technical work to be done here. Thus Deutsch  %%@
(2010) is a striking appeal to fellow-Everettians to stop defending the interpretation %%@
against accusations and rivals, and instead explore the new physics that it promises to %%@
contain; and in the course of this, he admits that the exploration will be very %%@
challenging, since no one has yet given a precise mathematical description of (even toy %%@
models of) this branching structure (2010, p. 546). (Earman's arguments, mentioned in %%@
Section \ref{collapse}, that various theorems make it hard to mesh the branching view with %%@
relativity, might be relevant here.)

Thus there are open questions about what more is required to articulate the Everettian %%@
vision. But however those further details go, the consequences for the philosophy of time %%@
will obviously be radical. A fundamentally deterministic (and relativistic) evolution of %%@
the universe's state will be ``overlaid'' by an approximate and emergent branching %%@
structure for time (and also by an apparent indeterminism). Again, I duck out of pursuing %%@
this issue.

But in closing: for  philosophers of time interested in the Everett interpretation, I must %%@
mention---and recommend---an analogy proposed by Saunders and Wallace between times as %%@
understood on the `block universe' or `B-theory' of time, and worlds or branches as %%@
understood by the Everettian. Recall that the B-theorist says:\\
\indent (i): reality is four-dimensional, and slices across it are in principle arbitrary %%@
and artefactual (especially in relativity theory, with no absolute simultaneity); but %%@
also\\
\indent (ii): for describing the history of the universe---and in particular, in physics, %%@
for doing dynamics---only a small subset of slices will be useful; though the criteria to %%@
select that subset will be a bit approximate.\\
Then the proposed analogy is as follows. Similarly, the Everettian says:\\
\indent (i'): the `slicing' of reality by choosing a basis in Hilbert space is in principle %%@
arbitrary and artefactual; but also\\
\indent (ii'): for describing the history of the universe---and in physics, for doing the %%@
approximate and emergent dynamics of a world---only a small subset of bases will be useful; %%@
though the criteria to select that subset will be a bit approximate (since decoherence %%@
gives no absolute criterion for a system-environment split, or for when interference terms %%@
are small enough).\footnote{For more details, cf. Wallace (2001); my own discussions are %%@
Butterfield (2001, Sections 7.3 and 8) and Butterfield (2011 Sections 3f.: a review of %%@
Saunders (2010)).}

\subsection{Time treated similarly in quantum and classical theories}\label{nodifference}
I turn to Section \ref{prosp}'s second deflating thought: that, interpretations aside, %%@
quantum theory's {\em formalism} has nothing special to say about time---it just tags along %%@
with whatever is said by the cousin classical theory, be it Newtonian or relativistic.

More precisely, let us distinguish three sorts of `cousin classical theory': Newtonian, %%@
special relativistic, and general relativistic. In the first two sorts of classical theory, %%@
there is a main common feature: viz. time is principally treated as a coordinate of %%@
spacetime (Newtonian or special relativistic), and as an independent variable or parameter %%@
in the equations of motion. Thus for example: the position of a particle, or the strength %%@
of an electric field, are functions of time, ${\bf q}(t), {\bf E}(t)$, with their %%@
time-evolution given by equations of motion. Agreed, we should note some qualifications of %%@
this statement.\\
 \indent (i): A field $\bf E$ is a function of space as well as time, so that classical %%@
field theories' equations of motion are partial, not ordinary, differential equations.\\
 \indent  (ii): Special relativity, with its frame-dependence of simultaneity, mingles %%@
space and time coordinates in ways that Newtonian physics does not.\\
 \indent (iii): Both Newtonian and special relativistic theories make occasional use of %%@
time as a function of other quantities. For example, they consider the time of arrival of a %%@
particle at a position $\bf x$, $t({\bf x})$; or they consider a clock, i.e. a system with %%@
a position  variable (or more generally, `indicator' variable), $q$ say, deliberately %%@
designed to be equal to the time: $q(t) \equiv t$.\\
 \indent (iv):  If one wishes, one can treat time in a manner more similar to the position %%@
of a particle. In both the Lagrangian and Hamiltonian formalism, $t$ is then treated as one %%@
of the configuration coordinates $q$, all of which are functions of a temporal parameter, %%@
$\tau$ say. I shall not go into details about this: Hilgevoord warns that it is liable to %%@
confuse time's roles as a parameter and as a dynamical variable (as in (iii): 1996, p. %%@
1452; 2002, p. 302; 2005, Section 2.5). For a detailed pedagogic treatment within classical mechanics, I recommend Johns (2005, Part II); and for two related approaches within quantum mechanics, I recommend Rovelli (2009), Reisenberger and Rovelli (2002), and Brunetti et al (2010).

The main point here is that, broadly speaking, quantum theories, be they Newtonian or %%@
special relativistic, tag along with what is said by these cousin classical theories; and %%@
even with the four qualifications. These quantum theories principally treat time as a %%@
coordinate, and as a parameter in the equations of motion: with appropriate qualifications %%@
to allow for field theories, and relativistic mingling of space and time coordinates. Here %%@
I should note some differences about qualification (iii), which will return in Section %%@
\ref{3times} below. In short: there are subtleties, and even controversies, about time as a %%@
physical quantity, because quantum theory represents quantities as operators, and there are %%@
subtleties about defining a time operator.\footnote{{Besides, the main point and its %%@
appropriate qualifications are not wholly independent of the interpretative controversies %%@
briefly surveyed in Section \ref{timeinterp}. For example, the pilot-wave
interpretation calls for special comments under (ii) and (iii). As to (ii), the %%@
interpretation tends not to be fundamentally relativistic. As to (iii), the fact that %%@
point-particles always have a position implies that the interpretation has no trouble in %%@
defining the time of arrival of a particle in a given spatial region, and similar notions %%@
like the time-period a particle spends in a region: notions which are hard to define in the %%@
orthodox formalism---cf. footnote 1.}\label{notrouble}}

When we turn to considering general relativity as the cousin classical theory, the %%@
situation becomes notoriously murky. There is no satisfactory corresponding quantum theory %%@
of gravity; nor even a consensus about what it would look like. And the stakes are high: %%@
finding such a theory is widely considered the holy grail of theoretical physics. (For %%@
philosophical introductions of the issues involved, cf. e.g. Butterfield and Isham 2001, %%@
Rovelli 2006, Huggett and Wuthrich (this volume).) So here the overall situation is like %%@
that in Section \ref{timeinterp}. A conclusive discussion of our topic, time in quantum %%@
physics, would have to wait till the search for such a theory had succeeded. And in the %%@
meantime, the discussion is liable to turn into a survey of the {\em pros} and {\em cons} %%@
of the various contenders---a survey not focussed on time.

But there is also a disanalogy with the situation  in Section \ref{timeinterp}. In the %%@
controversies about what a quantum theory of gravity should look like, one central and %%@
long-established theme has been the conflict between the way general relativity treats %%@
space and time, namely as dynamical (variable, and interacting with matter and radiation), %%@
and the way other theories, both classical and quantum, treat them, namely as non-dynamical %%@
(also called: `fixed' or `background'). Besides, in one main approach to quantum gravity, %%@
viz. the quantization of canonical (i.e. Hamiltonian) formulations of general relativity, %%@
this conflict is especially striking, indeed severe, as regards the treatment of time %%@
rather than space: and accordingly, it is called `the problem of time'. So this would be an %%@
appropriate focus for a discussion (admittedly inconclusive and controversial) of `time in %%@
quantum physics'. But in this paper, I shall not take this road.\footnote{For discussion of %%@
it, cf. Butterfield and Isham (1999, especially Sections 4 and 5), Earman (2002), Belot %%@
(2006, 2011) and Kiefer (2011).}

\subsection{Time-reversal: postponed to another day}\label{reverse}
Sections \ref{timeinterp} and \ref{nodifference} have listed several roads not taken.  Of %%@
course these  are yet others.  I shall mention---for an illusory sense of completeness, %%@
merely by allusion!---just one:  time-reversal.

 Broadly speaking, quantum theories have an operation of time-reversal on their space of %%@
states (often defined by analogy with a time-reversal operation on the classical %%@
state-space); and the theory's equations of motion (laws of evolution for the state) are %%@
time-reversal invariant---i.e. if the equations allow a certain evolution (temporal %%@
sequence of states, or possible history), then they also allow its time-reverse. But %%@
various questions arise: what is the best justification of the time-reversal operator's %%@
definition? And what should we make of experiments indicating that certain weak %%@
interactions violate time-reversal invariance? In particular, how does this bear on the %%@
direction or `arrow' (or: directions and arrows!), of time? For some comments in this %%@
volume on these questions, cf. Belot (this volume), Wallace (this volume).

I should also add that there are adjacent larger themes: (i) time-reversal in general, and %%@
in other specific theories such as electromagnetism (for which, cf. Earman (2002a), %%@
Malament (2004) and North (2008)); and (ii) the relations of time-reversal to other %%@
discrete symmetries, especially quantum field theory's CPT theorem  (for which, cf. Greaves %%@
(2010), Arntzenius (2011), Greaves and Thomas (2012).).

\section{Uncertainty principles in general}\label{UPgenl}
\subsection{Prospectus}\label{prosp2}
As I announced in Section \ref{prosp}, my main aim in the rest of this paper is to report %%@
the work of such authors as Busch, Hilgevoord and Uffink in sorting out long-standing %%@
confusions, especially about the time-energy uncertainty principle. This will include such %%@
topics as: distinguishing (following Busch) three roles for time (Section \ref{3times}); %%@
and time operators, in particular Pauli's influential `proof' that there cannot be one %%@
(Section \ref{opors}). So nothing that follows is original. And needless to say,  I will %%@
have to omit many details of these authors' work, let alone work by others. In particular, %%@
I will wholly exclude historical aspects of the topics I do treat: e.g.  the influence over %%@
the decades of the Bohr-Einstein photon box experiment, and of Pauli's proof; for which cf. %%@
e.g. Busch (1990, Section 3.1 and 4; 2008, Sections 2.3 and 6) and Hilgevoord (1998, 2005).

My overall theme will be that there are very different versions of the time-energy %%@
uncertainty principle, i.e. a relation like
\be
\Delta T \Delta E \geq \frac{1}{2}\hbar
\label{TE}
\ee
that are valid in different contexts. Some will use, not the variance, but rather some %%@
other measure of spread. Some will involve a time operator, but others will not; since, %%@
indeed, for some problems i.e. Hamiltonians, an appropriate time operator cannot exist.  %%@
(But so far, there is no established general theory of time measurements, telling us for %%@
which Hamiltonians a time operator exists.)

Furthermore: one of the main novelties in these uncertainty principles, viz. the use of %%@
measures of spread other than the variance, is just as important for quantities other than %%@
time and energy, such as position and momentum. So for clarity, this Section will report %%@
the definition of such measures, and how applying them leads to novel uncertainty %%@
principles for such quantities: I  postpone until Section \ref{3times} and %%@
\ref{UPintrinsic} their application to time-energy uncertainty principles. Here I will %%@
report two such measures of spread, and two corresponding uncertainty principles, in %%@
Sections \ref{UPwidth} and \ref{UPtranswidth} respectively.

But before doing so, I should say a little about the broader landscape. As I see matters, %%@
there are two main themes about uncertainty principles which I will ignore, but which are %%@
important and active research areas---indeed, at least as important as the work that I do %%@
discuss. The first theme concerns error and disturbance in measurement; the second concerns %%@
classical analogues.

\indent (1): {\em Error and disturbance}: The usual textbook form of the uncertainty %%@
principle (recalled at the start of Section \ref{UPwidth}) is naturally understood as a %%@
constraint on statistics, or (more or less equivalently) on state preparation. This is a limitation: not least because Heisenberg's original %%@
(1927) argument concerned how minimizing the error in a position measurement implied a %%@
greater disturbance in the system's momentum, and this argument spawned countless discussions of the %%@
uncertainty principle in terms of error and disturbance. (For an introduction to %%@
Heisenberg's thinking and its legacy, cf. e.g. Jammer (1974, Sections 3.2-3.5), Hilgevoord %%@
(2005, Section 3.2), Hilgevoord and Uffink (2006).) So one naturally asks whether there are %%@
rigorous uncertainty principles that stay more faithful to Heisenberg's own thinking, i.e. that are about a trade-off in errors and-or disturbances in measurements, especially in position and momentum. %%@
Indeed, such uncertainty principles have been developed, especially in the last twenty %%@
years, with modern methods for analysing quantum measurements, using such notions as operation, instrument and POVM on phase space. Thus Busch et al (2007) discuss three notions of error or inaccuracy in a measurement (Section 3.3), and derive three corresponding uncertainty principles for position and momentum (Section 4.3, eq. (45)-(47)). (Their derivations use uncertainty principles for width of the bulk, of the kind I will report in Section \ref{UPwidth}; cf. ibid., Section 2, eq. (9)-(13).) Another example: Ozawa (2003, 2004) defines an error $\varepsilon(A)$ of a measurement of a quantity $A$ and %%@
a disturbance $\eta(B)$ by the measurement on a quantity $B$, and then proves they are related by
\be
\varepsilon(A)\eta(B) + \varepsilon(A)\Delta(B) + \Delta(A)\eta(B) \geq  \frac{1}{2} | %%@
\langle [A,B] \rangle_{\rho} | \;\; ;
\label{OzawaUP}
\ee
where $\Delta$ denotes as usual the standard deviation in the state $\rho$. Besides, these studies bear on real experiments: a recent one is  Erhart et al. (2012) on neutron spin, and Busch et al (2007, Section 7) discuss many more.

\indent (2): {\em Classical analogues}: In recent decades, research in the symplectic %%@
geometry that underlies Hamiltonian classical mechanics has uncovered remarkable classical %%@
analogues of the uncertainty principle. The breakthrough was Gromov's 1985 symplectic %%@
no-squeezing theorem. It is a remarkable strengthening of Liouville's theorem, familiar %%@
from textbooks of mechanics, that any canonical transformation is volume-preserving. For it %%@
says, roughly speaking, that the surface area, defined by the projection of a phase space %%@
region onto the plane given by conjugate  coordinates $q_i$ and $p_i$, is also preserved. %%@
In the light of the Gospel (`It is easier for a camel to go through the eye of a needle, %%@
than for a rich man to enter into the kingdom of God' (Mark, 10:25)), this theorem has been %%@
nicknamed `the symplectic camel'. Thus the theorem says: a ball in Hamiltonian phase space %%@
of radius $R$ (`the camel') can be canonically transformed so as to pass through an `eye of %%@
a needle' defined by a disc of radius $r$ in a conjugate coordinate plane $(q_i, p_i)$ {\em %%@
only if} $r > R$.  For a glimpse of this rich subject, and its application to the study of %%@
the classical-quantum relationship, cf. De Gosson (2001, Section 3.7; 2009), De Gosson and %%@
Luef (2009).

So much by way of glimpsing the broader landscape---and seeing that uncertainty principles %%@
remain a rich field for research in both physics and philosophy.

\subsection{An uncertainty principle for width of the bulk}\label{UPwidth}
We begin by recalling the usual textbook `Heisenberg-Robertson' uncertainty principle (e.g. %%@
Jauch 1968, p. 161; Isham 1995, Section 7.3): for any quantities $A, B$, and quantum state %%@
(density matrix) $\rho$:
\be
\Delta _{\rho}A \Delta _{\rho}B \geq  \frac{1}{2} | \langle [A,B] \rangle_{\rho} | \;\; ;
\label{RobUP}
\ee
with the special case, from $[Q,P] = i \hbar {\bf I}$:
\be
\Delta _{\rho}Q \Delta _{\rho}P \geq \frac{1}{2}\hbar \; .
\label{HU}
\ee
This uncertainty principle has various kinds of limitation. As mentioned in (1) of Section %%@
\ref{prosp2}, one is that it cannot be interpreted in terms of a balance between a %%@
measurement's error or `noise' in its outcome, and its disturbance of another quantity---as %%@
Heisenberg originally intended. Another kind of limitation is that the right-hand side of %%@
eq. \ref{RobUP} is state-dependent and so may vanish (for example if $\rho$ is $| \psi %%@
\rangle \langle \psi |$ for $| \psi \rangle$ an eigenvector of $A$ or $B$), and so fail to %%@
provide a non-trivial lower bound on the left-hand side. But we  will concentrate on a %%@
third kind of limitation, concerning {\em measures of spread}.

In this regard, the first points to make  are that the standard deviation of some
quantum states diverges; and relatedly, that it can be very large for states that are %%@
intuitively `well-concentrated'. A standard example is the Breit-Wigner state, whose %%@
probability distribution for energy is the Cauchy distribution: this has a parameter %%@
$\gamma$, and becomes arbitrarily concentrated around 0 as $\gamma \rightarrow 0$---but has %%@
an infinite standard deviation for all $\gamma > 0$ (e.g. Hilgevoord and Uffink 1988, p. %%@
99; Uffink 1990, p. 120). In this sense, the traditional eq.s \ref{RobUP} and \ref{HU} are %%@
logically weak: they may not apply because the standard deviation diverges, and if they do apply, they allow the state to be concentrated in both the quantities, e.g. in both $Q$ and $P$.\footnote{Thus this
limitation lies in the ``opposite direction'' to that in (1) of Section \ref{prosp2}: which %%@
was that eq.s \ref{RobUP} and \ref{HU}, if {\em mis}-interpreted in terms of error and %%@
disturbance, are too strong---indeed, they have  been experimentally violated.}

This situation prompts one to define other measures of spread than the standard deviation %%@
(or variance).
And indeed: some of these measures will not require that the quantity concerned is %%@
represented by a corresponding operator.

One way to measure the spread of a probability distribution is along the following lines: %%@
the length $W_{\alpha}$ of the smallest interval on which a sizeable fraction $\al$ of %%@
distribution is supported. (Here, `sizeable' can be taken to mean $\al \geq \frac{1}{2}$; %%@
and we will not need to worry about there being two smallest intervals of equal size.). %%@
That is: we represent the spread of a distribution in terms of the smallest interval on %%@
which the bulk of the distribution is found. We call this a {\em width}, or {\em width of %%@
the bulk}; (it is also sometimes called `overall width', to distinguish it from a %%@
`translation width' to be introduced in Section \ref{UPtranswidth}).

It can then be shown that for any quantum state, the widths of the position and momentum %%@
distributions in one spatial dimension satisfy:
\be
W_{\alpha}(Q) W_{\alpha}(P) \geq c_{\al}\hbar \;\; \mbox{if } \al \geq \frac{1}{2};  \;\; %%@
\mbox{with }  c_{\al} \;\; \mbox{of order } 1; \mbox{ namely, } c_{\al} = 2\pi(2 \al - %%@
1)^2;
\label{UU}
\ee
(Hilgevoord and Uffink 1990, p. 129; Uffink 1990, Sections 2.4.3, 2.5.4; based on results %%@
by Landau and Pollak (1961)).

Two comments. (1): As stated, this result depends on the position and momentum %%@
representations being Fourier transforms. Thus we can state it a bit more generally and %%@
explicitly, in terms of Fourier transforms; as follows. For any normalized $L^2$ function %%@
i.e. $f$ such that $\int | f |^2 dx = 1$, we define $W_{\al}(|f|^2)$ to be the width of the %%@
smallest interval $J$ such that $\int_J | f |^2 dx = \al$; and similarly for the Fourier %%@
transform $\hat f$. Then the result is:
\be
W_{\al}( |f|^2 )W_{\al}( |{\hat f}|^2 ) \geq c_{\al}\hbar \;\; \mbox{if } \al \geq %%@
\frac{1}{2};  \;\; \mbox{with }  c_{\al} = 2\pi(2 \al - 1)^2.
\label{UUFourier}
\ee
(2): In fact, this result arises from more general results based on  the {\em statistical %%@
distance} between two probability distributions: an idea which makes sense for discrete %%@
(and so: not Fourier-related) distributions, and which has wide application. In particular, %%@
this idea leads to {\em entropic} uncertainty relations, and Uffink's generalizations of %%@
them using his $M_r$ measures, given by his representation theorem for measures of %%@
uncertainty (Uffink 1990, Sections 1.5.2, 2.4.4, 3.5.6-7; Maassen and Uffink (1988)). But I %%@
cannot go into detail: suffice it to say that statistical distance also underlies the idea %%@
of translation width, introduced in the next Section.

\subsection{An uncertainty principle for translation width, and width of the %%@
bulk}\label{UPtranswidth}
Both the measures of uncertainty so far introduced---the familiar notion of variance, and %%@
Section \ref{UPwidth}'s notion of  width---are about uncertainty in the value of a %%@
quantity. But one can instead (and: also!) be uncertain about the system's {\em state}. %%@
There is a natural measure of this second kind of uncertainty; and Hilgevoord and Uffink %%@
show the remarkable result that it combines with Section \ref{UPwidth}'s notion of the %%@
width $W_{\al}$, to give uncertainty principles. Thus these uncertainty principles involve %%@
two different notions of spread, each of them different from the familiar notion of %%@
variance. They also provide an exactly similar treatment of space and time---and thus also %%@
of the respective conjugate notions, energy and momentum. But I will postpone the %%@
discussion of time until Section \ref{3times} onwards. So this Subsection is by way of an %%@
appetizer for later, especially Section \ref{HUtransbulk}.

Consider the task of trying to distinguish a state  $| \phi \rangle$ from another $| \psi %%@
\rangle$. Intuitively, this is easier the closer the state are to being orthogonal. So %%@
given $|\langle \phi |  \psi \rangle | = 1 - r$, with $0 \leq r \leq 1$, we will call $r$ %%@
the {\em reliability} with which $| \phi \rangle$ and $| \psi \rangle$ can be %%@
distinguished. So if $| \phi \rangle = | \psi \rangle$, then $r = 0$; while if $| \phi %%@
\rangle$ and $| \psi \rangle$ are orthogonal, $r$ attains its maximum value, 1.

We now apply reliability to the translation of a quantum state $| \psi \rangle$ in space %%@
(in one spatial dimension, which we now label $x$). Translation is effected by the %%@
exponentiation of  the total momentum, i.e. by the unitary operators:
\be
 U_x(\xi) = \exp(-i P_x \xi / \hbar) \; .
\label{transUp}
\ee
Then for given $r \in [0,1]$, we define $\xi_{r}$ as the smallest distance for which
\be
|\langle \psi | U_x(\xi_{r}) | \psi \rangle | = 1 - r \;.
\ee
Following Hilgevoord and Uffink (1988, p. 103; 1990, p. 125), we will call $\xi_{r}$ the %%@
{\em spatial translation width} of the state $| \psi \rangle$.

As mentioned at the end of Section \ref{UPwidth}, translation width is closely connected to %%@
the idea of statistical distance. But I cannot go into details (cf. Hilgevoord and Uffink %%@
1991). For this paper, all I need is to note the crucial distinction:  ${\xi_{r}}$ for a %%@
given (pure) state $| \psi \rangle$ is by no means the same as the spatial width %%@
$W_{\al}(q)$ , i.e. the width of the probability distribution $| \langle q | \psi \rangle %%@
|^2$ of the position operator $\hat Q$ for the state  $| \psi \rangle$. Agreed, in `simple' %%@
cases, i.e. if  $| \langle q | \psi \rangle |^2$ has a single peak, the two measures will %%@
be close. In such a case: if intuitively the bulk of the distribution is on an interval of %%@
length $d$, then $W_{\al}$, for $\al$ close to 1, will be
close to $d$; and ${\xi_{r}}$, for $r$ close to 1, will also be close to $d$. But suppose %%@
$| \langle q | \psi \rangle |^2$ has many narrow peaks each with a small and similar width, %%@
of about $e$ say,
while the entire distribution is spread over a much larger interval $d$. (Of course, %%@
interference patterns provide paradigm examples of this: and indeed, the translation width %%@
crops up  in optics, albeit under other names like `resolving power' and `Rayleigh's %%@
criterion [for distinguishing diffraction patterns]'.)  Then ${\xi_{r}}$ will be of the %%@
order $e$, since a translation by a mere $e$ will suffice to move the peaks into the %%@
troughs and {\em vice versa}, securing distinguishability (near-orthogonality);  while %%@
$W_{\al}(q)$ will be of order $d$.

The translation width $\xi$ combines with Section \ref{UPwidth}'s width of the bulk $W$ to %%@
give  uncertainty principles.  To prepare for a later discussion, we will define the  width %%@
of the bulk for momentum more explicitly than we did in eq.s \ref{UU} and \ref{UUFourier}. %%@
Thus let  $| p_x \rangle$ denote a complete set of eigenstates of $P_x$. (We set aside %%@
degeneracy, for a simpler notation.) So, with the integration symbol perhaps including a %%@
sum over discrete eigenstates, we have
\be
 \int | p_x \rangle \langle p_x | dp_x \; = {\bf I} \; .
\label{residy1}
\ee
Then we define the width  $W_{\al}(P_x)$ of the momentum distribution as the smallest %%@
interval such that
\be
\int_{W_{\al}(P_x)} | \langle p_x | \psi \rangle |^2 dp_x = \al \;.
\ee
Then it can be shown (Uffink and Hilgevoord 1985, Appendix D; Hilgevoord and Uffink 1988, %%@
pp. 103-105; 1990, p. 134) that for $r \geq 2(1- \al)$:
\be
 \xi_{r} W_{\al}(P_x) \geq  C(\al, r) \hbar \; ;
\label{GenUU1}
\ee
where for sensible values of the parameters, say $\al = 0.9$ or 0.8, and $0.5 \leq r \leq %%@
1$, the constant $C(\al, r)$ is of order 1. More precisely: $C(\al, r) = 2 \arccos \frac{2 %%@
- r - \al}{\al}$ for $r > 2(1 - \al)$.

\subsubsection{The principle's significance}\label{13Comments}
Hilgevoord and Uffink give judicious discussions of the merits and significance of eq. %%@
\ref{GenUU1} (Uffink and Hilgevoord (1985, p. 938); Hilgevoord and Uffink (1988, pp. %%@
105-108; 1990, p. 134); Hilgevoord (1998, Section 4); Hilgevoord and Atkinson (2011, %%@
Section 4)). In particular, they show how it captures Heisenberg's thinking in his original %%@
microscope argument for the uncertainty principle; (cf. (1) in Section \ref{prosp2})). I %%@
shall just report four of their main comments. The first two are formal; the third and %%@
fourth more physical.

First: the inequality is completely general, in the senses that (i) its right-hand side %%@
provides a state-independent lower bound, and (ii) it depends only on the existence of %%@
translation operators eq. \ref{transUp} and the completeness relations, eq. \ref{residy1}. %%@
Second: thanks to (ii), the equation is relativistically valid.

Third: broadly speaking eq. \ref{GenUU1} gives more information than the traditional %%@
Heisenberg-Robertson uncertainty principle, eq. \ref{HU}. It is not just that as mentioned %%@
in Section \ref{UPwidth}, a state may not have a standard deviation. Also, there are two %%@
other points.\\
 \indent \indent (a): We saw that ${\xi_{r}}$ should be distinguished from the spatial %%@
width $W_{\al}(q)$: in an interference pattern we can have ${\xi_{r}} << W_{\al}(q)$. In %%@
such a case, eq \ref{GenUU1} is stronger than eq. \ref{HU}; and even than eq. \ref{UU}.\\
 \indent \indent  (b): Eq. \ref{GenUU1} shows that in a many-particle system the width of %%@
the {\em total} momentum can control whether the positions of {\em each} of the component %%@
particles can be sharply determined.  For if the spread of just one particle position is %%@
small, then ${\xi_{r}}$ for the entire system's state is small. (For distinguishing %%@
multi-component states under
translation only requires distinguishing one component: recall that orthogonality in one %%@
factor of a tensor product implies orthogonality in the whole.)  And so, by eq. %%@
\ref{GenUU1}, the width in the total momentum must be large. Conversely, if $W_{\al}(P_x)$ %%@
is small, ${\xi_{r}}$ must be large; and so the spread of each component particle's position must be large.

Finally, the fourth comment looks ahead to the topic of time. Namely: in Section %%@
\ref{HUtransbulk}, we will see an exactly parallel treatment for time. There will be an %%@
uncertainty principle combining a temporal translation width with the bulk-width of the %%@
energy distribution.

\section{Three roles for time}\label{3times}
Busch (1990, Section 2; 2008 Section 2) articulates a trichotomy of roles (or senses) of %%@
time, which is invaluable for organizing discussion (i.e. preventing confusion!) about time %%@
in quantum physics. Indeed, it will be clear that it is equally useful for discussing time %%@
in classical physics; (albeit with some obvious adjustments, e.g. adjusting the %%@
representation of physical quantities from self-adjoint operators, to functions on a phase %%@
space). I shall present each role in a Subsection.  Beware: Busch's trichotomy, and his %%@
labels for the three roles, is not used by most other authors; (and he suggests more than %%@
one label for each role---I shall adopt the labels of his later discussion (2008)).

Although I follow Busch, it will be clear that the trichotomy, or rather the three binary %%@
distinctions it involves, are common currency: they are accepted by other authors, although %%@
under a different label (or none). For example, the distinction between Busch's first role, %%@
and his second and third roles is in effect the distinction which I stated at the start of %%@
Section \ref{nodifference}, as common to classical and quantum physics: namely, between (a) %%@
time as a coordinate of spacetime and an independent variable in the equations of motion, %%@
and (b) time as a function of other quantities, as in the pointer-variable of a clock (cf. %%@
(iii) of Section \ref{nodifference}).

This distinction is also emphasized by Hilgevoord (1996 Section 2; 2002, Section 2; 2005, %%@
Sections 1, 2; Hilgevoord and Atkinson 2011, Section 2). He argues persuasively that much %%@
of the historical confusion about time in quantum theory (including such figures as von %%@
Neumann and Pauli) can be dispelled by respecting this distinction, and the corresponding %%@
distinction about space, or position: viz. between (a) position as a coordinate of %%@
spacetime, and (b) position as a dynamical variable, e.g. the position of a point-particle. %%@
In particular, Hilgevoord suggests that:\\
\indent (i): for clarity, we should write $x$ for the first, and $q$ for the second, and so %%@
write a wave-function as $\psi(q)$, not $\psi(x)$; \\
\indent (ii): armed with these distinctions, there is no temptation to say that quantum %%@
physics marks a break from classical physics, by making position but not time into a %%@
dynamical variable. On the contrary: both classical and quantum physics treat both time and %%@
position, either as a coordinate, or as a dynamical variable.

\subsection{External time}\label{external}
Busch suggests `external time' (or in his 1990: `pragmatic time') for time as measured by %%@
clocks that are not coupled to the objects studied in the experiment. So in this role, time
specifies a parameter or parameters of the experiment: e.g. an instant or duration of %%@
preparation or of measurement, or the time-interval between preparation and measurement. In %%@
this role, there thus seems to be no scope for uncertainty about time.

But, as Busch discusses, there is tradition (deriving from the founding fathers of quantum %%@
theory) of an uncertainty principle between (i) the {\em duration} of an energy %%@
measurement, and (ii) {\em either} the range of an uncontrollable change of the measured %%@
system's energy {\em or} the resolution of the energy measurement {\em or} the statistical %%@
spread of the system's energy. Busch argues, and I agree, that Aharonov and Bohm (1961) %%@
refute this tradition; (Busch 1990a, Section 4; 2008 Section 3.1). They give a simple model %%@
of an arbitrarily accurate and arbitrarily rapid energy measurement. In short: two %%@
particles are confined to a line and are both free, except for an impulsive measurement of %%@
the momentum and so energy of the first by the second, with the momentum of the second %%@
being the pointer-quantity. I will not go into detail: but it is worth noting, following %%@
Busch, that a proper analysis and vindication of Aharonov and Bohm's refutation uses POVMs: %%@
a generalized notion of physical quantity developed in recent decades.

\subsection{Intrinsic times}\label{intrinsic}
Busch suggests `intrinsic time' (or in his 1990: `dynamical time') for a dynamical variable %%@
of the studied system, that functions to measure the time. For example: the position of a %%@
clock's dial relative to the background, i.e. the clock's face.  Another example---very %%@
much Pickwickian or `in principle'---is the position of a classical free particle: it also %%@
registers the time, if only one could (i) `read off' its position relative to the %%@
background space and (ii) keep up with its motion, so as to be able to read off that %%@
position! Busch suggests, as does this Pickwickian example, that in principle every %%@
non-stationary quantity $A$ defines for any quantum state $\rho$ a characteristic time %%@
$\tau_{\rho}(A)$ in which $\langle A \rangle$ changes `significantly'. For example: if $A = %%@
Q$, and $\rho$ is a wave packet, say $\rho = | \psi \rangle \langle \psi |$, then %%@
$\tau_{\rho}(A)$ could be defined as the time for the bulk of the wave packet to shift by %%@
its width, in the sense of either Section \ref{UPtranswidth} or Section \ref{UPwidth}.

This suggests we should expect there will be various uncertainty principles for various %%@
definitions of intrinsic times. Indeed this will be so: details in Section %%@
\ref{UPintrinsic}. But first, we should note that in some of the imaginable experiments, or %%@
more generally scenarios, in which a dynamical variable registers a time, the time %%@
concerned has a physical meaning `as a time': for example, the time of arrival of a %%@
particle in a spatial region. This leads to Busch's third role for time ...

\subsection{Observable time: time operators}\label{opors}
Busch suggests `observable time' (also called, in his 1990: `event time') for the sort of %%@
case just mentioned, where the measured time has a physical meaning `as a time': e.g. a %%@
time of arrival, or of sojourn, in a given spatial region, or a time of flight between two %%@
given places, or a time of decay.\footnote{So here we return to the examples that, as %%@
mentioned in footnote \ref{notrouble}, are for the pilot-wave interpretation a case of `no %%@
worries', since its particles always have a definite position.}

In quantum theory, such cases prompt the idea of a time operator, i.e. a self-adjoint %%@
operator whose spectrum is the possible values of the time in question: (hence the title of %%@
this Section). And so here we meet  Pauli's one-liner `proof' that there cannot be such an %%@
operator. As I said, I will not go into the history of this proof and its legacy: not least %%@
because it is so well covered by Busch and Hilgevoord. I will just mention what I take to %%@
be the two main points: the first `for Pauli', the second `against him'.

(1): The first point expresses Pauli's idea that a time operator would imply that the %%@
Hamiltonian has the entire real line as its spectrum: forbidding discrete eigenvalues, and %%@
also making the Hamiltonian unphysical because unbounded below. Thus: a self-adjoint %%@
operator $T$ generating translations in energy according to
\be
\exp(i \tau T / \hbar) H \exp(-i \tau T / \hbar) = H + \tau {\bf I} \; , \;\;\; \forall %%@
\tau \in \mathR
\label{CovHT}
\ee
would imply that the spectrum of $H$  is $\mathR$. The reason is simply that unitary transformations are spectrum-preserving, i.e. the left-hand side has the same spectrum as the $H$. So $H$ has the same spectrum as $H + \tau {\bf I}$ for all $\tau$; so the spectrum of $H$ is $\mathR$.

Here we should note that eq. \ref{CovHT} would imply that in a dense domain:
\be
[H,T] = i \hbar {\bf I} \; .
\label{CCRHT}
\ee
And this would imply (cf. eq \ref{RobUP}) our opening `prototype form' of the time-energy %%@
uncertainty principle for any state $\rho$
\be
\Delta_{\rho} T \Delta_{\rho} H \geq \frac{1}{2}\hbar \; .
\label{THrho}
\ee

(2): But the converse fails. That is: eq \ref{CCRHT} does {\em not} imply eq. \ref{CovHT}. %%@
So the door is open for some realistic Hamiltonians (i.e. bounded below and-or with %%@
discrete spectrum; i.e. without the whole of $\mathR$ as spectrum), to have a $T$ %%@
satisfying  eq \ref{CCRHT}---which Busch (2008) calls a {\em canonical time operator}. %%@
Busch points out that although there is little general theory of which Hamiltonians have %%@
such a $T$ (even when we generalize the concept of quantity to POVMs), there are plenty of %%@
examples---including familiar systems like the harmonic oscillator. For details, I refer to %%@
his discussion (Busch 2008, Section 6), and to other related results (Galapon 2002, %%@
Hegerfeldt and Muga 2010).

\section{Time-energy uncertainty principles with intrinsic times}\label{UPintrinsic}
We saw at the end of Section \ref{intrinsic} that we should expect various such principles, %%@
with various intrinsic times (represented by a parameter, not an operator!). And indeed %%@
there are: details of several such principles can be found Busch (1990, Section 3.3; 2008, %%@
Section 4). I shall present just:\\
\indent (i) one well-known principle, due to Mandelstam and Tamm (Section \ref{MandTamm}); %%@
and\\
\indent (ii) a principle due to Hilgevoord and Uffink, which gives a parallel treatment for %%@
time and energy, to that in Section \ref{UPtranswidth} for position and momentum (Section %%@
\ref{HUtransbulk}).

\subsection{Mandelstam-Tamm uncertainty principle}\label{MandTamm}
Mandelstam and Tamm combine the Heisenberg equation of motion of an arbitrary quantity $A$
\be
i \hbar \frac{dA}{dt} = [A,H]
\label{Heqmot}
\ee
with the Heisenberg-Robertson uncertainty principle, eq. \ref{RobUP}, and the definition of %%@
a characteristic time
\be
\tau_{\rho}(A) \; := \; \Delta_{\rho} A \; / \; | \; d \langle A \rangle_{\rho} \; / dt \; %%@
|
\label{tauMT}
\ee
and deduce
\be
 \tau_{\rho}(A) \Delta_{\rho}(H) \geq  \frac{1}{2} \hbar .
\label{MTUP}
\ee
Eq. \ref{MTUP} has many applications: (setting aside, now, my earlier scepticism about the %%@
standard deviation as a measure of spread). I shall report two examples.

Example 1: a free particle in a pure state $\psi$ with a sharp momentum i.e. %%@
$\Delta_{\psi}P << | \langle P \rangle_{\psi} |$. We can deduce that $\Delta_{\psi} Q(t) %%@
\approx \Delta_{\psi} Q(t_0)$, i.e. slow spreading of the wave-packet; so that the %%@
Mandelstam-Tamm time $\tau_{\rho}(Q)$, as defined by eq \ref{tauMT}, is indeed the time it %%@
takes for the packet to propagate a distance equal to its standard-deviation.

Example 2: the lifetime of a property $\hat P$. Define $p(t) := \langle \psi_0 | U^{-1}_t %%@
{\hat P} U_t \psi_0 \rangle$, with of course $U_t := \exp(-itH/\hbar)$. Then the %%@
Mandelstam-Tamm uncertainty relation gives
\be
| \frac{dp}{dt} | \leq \frac{2}{\hbar} (\Delta_{\psi_0} H) [p (1 - p)]^{\frac{1}{2}} \; .
\ee
Integration with initial condition $p(0) \equiv 1$ (i.e. $\hat P$ actual at $t = 0$) yields
\be
p(t) \geq \cos^2 (t(\Delta_{\psi_0} H) \; /\; \hbar) \;\; \mbox{for} \;\; 0 < t < %%@
\frac{\pi}{2}\frac{\hbar}{(\Delta_{\psi_0} H)}
\ee
so that if we define the lifetime $\tau_{\hat P}$ of the property $\hat P$ by $p(\tau_{\hat %%@
P}) := \frac{1}{2}$, we deduce
\be
\tau_{\hat P}.\Delta_{\psi_0} H \geq \frac{\pi \hbar}{4}.
\label{MTlifetime}
\ee
The general point is that the rate of change of a property of the system decreases with %%@
increasing sharpness of the energy: and in the limit of an energy eigenstate, of course, %%@
all quantities have stationary distributions.

\subsection{The Hilgevoord-Uffink `translation-bulk' uncertainty %%@
principle}\label{HUtransbulk}
Finally, I report (following Hilgevoord 1996, Section 4; Hilgevoord and Atkinson 2011, %%@
Section 4) how to combine:\\
\indent (i) use of $W_{\al}$ as in Section \ref{UPwidth} (cf. eq. \ref{UU}); with\\
\indent (ii) use translations-widths as in Section \ref{UPtranswidth} (cf. eq. %%@
\ref{GenUU1}); with \\
\indent (iii) similar treatments of space and time.

We apply the idea of reliability (discrimination of states) to the translation of a quantum %%@
state $| \psi \rangle$ in time {\em instead of} in space. Translation is effected by  %%@
exponentiation of the energy, i.e. by the unitary operators:
\be
U_t(\tau) = \exp(-iH \tau / \hbar) \;\; .
\label{transU}
\ee
Then for given $r \in [0,1]$, we define $\tau_{r}$ as the smallest time for which
\be
|\langle \psi | U_t(\tau_{r}) | \psi \rangle | = 1 - r \;.
\ee
$\tau_{r}$ may be called the {\em temporal translation width} of the state $| \psi %%@
\rangle$. If we choose $r$ such that $(1 - r)^2 = \frac{1}{2}$, i.e. $r = 1 - %%@
\surd{\frac{1}{2}}$, then $\tau_{r}$ is the {\em half-life} of the state $| \psi \rangle$; %%@
(cf. example 2 in Section \ref{MandTamm}).

The translation widths $\tau_{r}$, and $\xi_{r}$ from Section \ref{UPtranswidth}, combine %%@
with the widths $W$ introduced previously (cf. eq. \ref{UU}) in uncertainty relations. To %%@
avoid referring back, and yet bring out the strong analogy between position-momentum and %%@
time-energy, I will repeat the details for position-momentum, from Section %%@
\ref{UPtranswidth}.

In order to define the widths $W$ for energy and momentum, let $| E \rangle$ and $| p_x %%@
\rangle$ denote complete sets of eigenstates of $H$ and $P_x$ respectively. (We again set %%@
aside degeneracy, for a simpler notation.) So, with integration perhaps including a sum %%@
over discrete eigenstates
\be
\int | E \rangle \langle E | dE = {\bf I} \;\;\; \mbox{and} \;\;\; \int | p_x \rangle %%@
\langle p_x | dp_x \; = {\bf I} \; .
\label{residy}
\ee
Then we define the widths $W_{\al}(E)$ and $W_{\al}(P_x)$ of the energy and momentum %%@
distributions as the smallest intervals such that
\be
\int_{W_{\al}(E)} | \langle E | \psi \rangle |^2 dE = \al \;\;\; ; \;\;\;
\int_{W_{\al}(P_x)} | \langle p_x | \psi \rangle |^2 dp_x = \al \;.
\ee
Then it can be shown  (Uffink and Hilgevoord 1985, Appendix D; Hilgevoord and Uffink 1988, %%@
pp. 103-105; 1990, p. 134) that for $r \geq 2(1- \al)$:
\be
\tau_{r} W_{\al}(E) \geq  C(\al, r) \hbar \;\;\; , \;\;\; \xi_{r} W_{\al}(P_x) \geq  C(\al, %%@
r) \hbar \;\; ;
\label{GenUU}
\ee
where for sensible values of the parameters, say $\al = 0.9$ or 0.8, and $0.5 \leq r \leq %%@
1$, the constant $C(\al, r)$ is of order 1. More precisely: $C(\al, r) = 2 \arccos \frac{2 %%@
- r - \al}{\al}$ for $r > 2(1 - \al)$.

The comments from Section \ref{13Comments}---both formal comments, and comments about these %%@
inequalities' physical significance---carry over. I shall not repeat them, but just note %%@
their gist: that  eq. \ref{GenUU} gives more information than the traditional %%@
Heisenberg-Robertson uncertainty principle, like eq. \ref{HU}. In particular, comment (b) %%@
in Section \ref{13Comments}, about the width of the {\em total} momentum controlling the %%@
width of position for {\em each} of the component particles, carries over to energy and time. That is: %%@
eq. \ref{GenUU} implies that in a many-particle system, the width in the total energy %%@
can control whether the temporal spread, in the sense of temporal translation width, of {\em each} component particle is small. Thus if $W_{\al}(E)$ is small, $\tau_r$ must be large; and so the temporal spread of each component particle must be large.

To conclude: I hope to have shown that nowadays there is still plenty to study and to explore, both for physics and philosophy, about the roles of time in quantum physics, and in particular about the uncertainty principle---or better, {\em principles}.\\

{\em Acknowledgements}:--- I am grateful to: Paul Busch, Jan Hilgevoord and Jos Uffink for %%@
the inspiration of their writings and of their talks; to them and Matthew Donald, Tom Pashby, Bryan Roberts, David Wallace, Daniel Wohlfarth and James Yearsley for %%@
comments on earlier versions; to audiences in Traunkirchen, Cambridge and Dubrovnik; and to the editors for their patience. For hospitality, I am very grateful to Professor Anton %%@
Zeilinger and his colleagues in Vienna; and for a grant supporting my visit to Professor %%@
Zeilinger's group, I am grateful to the John Templeton Foundation. This work is also %%@
supported in part by a grant from the Templeton World Charity Foundation, whose support is %%@
gratefully acknowledged.

\section{References}

Aharonov, Y. and Bohm, D. (1961), `Time in the quantum theory and the uncertainty relation %%@
for time and energy', {\em Physical Review} {\bf 122}, p. 1649.

Arntzenius, F. (2011), `The CPT theorem', in Callender (2011), pp. 633-646.

Bedingham, D. et al. (2011), `Matter Density and Relativistic Models of Wave Function %%@
Collapse', arXiv:1111.1425.

Bell, J. (1986), `Six possible worlds of quantum mechanics', {\em Proceedings of the Nobel %%@
Symposium 65} (Stockholm August 1986); reprinted in Bell (1987), page references to %%@
reprint.

Bell, J. (1987), {\em Speakable and Unspeakable in Quantum Mechanics}, Cambridge: Cambridge %%@
University Press; second edition 2004, with an introduction by Alain Aspect.

Belot, G. (2006), `The representation of time in mechanics', in J. Butterfield and J. %%@
Earman (eds.), {\em Philosophy of Physics}, in the series {\em The Handbook of Philosophy %%@
of Science}, Amsterdam: Elsevier, pp. 133-227.

Belot, G. (2011), `Background independence', {\em General Relativity and Gravitation} {\bf %%@
43}, pp. 2865-2884

Belot, G. (this volume), `Temporal aspects of relativistic spacetimes'.

Bohm, D. and Hiley B. (1992), {\em The Undivided Universe}, London: Routledge.

Broad, C. (1923), {\em Scientific Thought}, New York: Harcourt, Brace and Co.

Brunetti, R., Fredenhagen, K. and Hoge M. (2010), `Time in quantum physics: from an external parameter to an intrinsic observable', {\em Foundations of Physics} {\bf 40}, pp. 1368-1378.

Busch, P. (1990),  `On the energy-time uncertainty relation: Part I: Dynamical time and %%@
time indeterminacy', {\em Foundations of Physics} {\bf 20}, pp. 1-32.

Busch, P. (1990a),  `On the energy-time uncertainty relation: Part II: Pragmatic time %%@
versus energy indeterminacy', {\em Foundations of Physics} {\bf 20}, pp.  33-43.

Busch, P. (2008), `The time-energy uncertainty relation', Chapter 3 of {\em Time in Quantum %%@
Mechanics}, ed. G. Muga et al. Springer Verlag (second edition, 2008); quant-ph/0105049.

Busch, P., Heinonen T. and Lahti P. (2007), `Heisenberg's uncertainty principle', {\em  Physics Reports} {\bf 452}, pp.  155-176.

Butterfield, J. (2001), `Some Worlds of Quantum Theory ', in R.Russell, J. Polkinghorne et %%@
al (ed.), {\em Quantum Mechanics (Scientific Perspectives on Divine Action vol 5)}, Vatican %%@
Observatory Publications, 2002; pp. 111-140. Available online at: %%@
arXiv.org/quant-ph/0105052; and at http://philsci-archive.pitt.edu/archive/00000204.

 Butterfield, J. (2007), `Reconsidering relativistic causality', {\em International Studies %%@
in Philosophy of Science} {\bf 21}, pp. 295-328: available at: %%@
http://uk.arxiv.org/abs/0708.2189; and at:
 http://philsci-archive.pitt.edu/archive/00003469/

Butterfield, J. (2011), Critical Notice of Saunders et al (eds.) (2010), {\em Philosophy}, %%@
{\bf 86}, pp. 451-463.

Butterfield, J. and Isham C. (1999), `On the emergence of time in quantum gravity', in J. %%@
Butterfield ed. {\em The Arguments of Time}, The British Academy and Oxford University %%@
Press, pp. 111-168: available at: gr-qc/9901024; and \\ %%@
http://philsci-archive.pitt.edu/archive/00001914/

Butterfield, J. and Isham C. (2001), `Spacetime and the philosophical challenge of quantum %%@
gravity', in C. Callender and N. Huggett (ed.s), {\em Physics Meets Philosophy at the %%@
Planck Scale}, Cambridge University Press, pp. 33-89: available at:  gr-qc/9903072; and  %%@
philsci-archive.pitt.edu/archive/00001915/

Callender, C. ed. (2011), {\em The Oxford Handbook of Philosophy of Time}, Oxford %%@
University Press.

De Gosson, M. (2001), {\em The Principles of Newtonian and Quantum Mechanics}, Imperial %%@
College Press.

De Gosson, M. (2009), `The symplectic camel and the uncertainity principle: the tip of an %%@
iceberg?', {\em Foundations of Physics} {\bf 39}, pp. 194-214.

De Gosson, M.  and Luef, F. (2009), `Symplectic capacities and the geometry of uncertainty: %%@
the irruption of symplectic topology in classical and quantum mechanics', {\em Physics %%@
Reports} {\bf 484}, pp. 131-179.

Deutsch, D. (2010), `Apart from universes', in Saunders et al (eds.) (2010), pp. 542-552.

Earman, J. (2002), `Thoroughly modern McTaggart: or what McTaggart would have said if he %%@
had read the general theory of relativity', {\em Philosophers Imprint} {\bf 2}, pp. 1-28.

Earman, J. (2002a), `What time reversal invariance is  and why it matters', {\em %%@
International Studies in the Philosophy of Science}, {\bf 16}, pp. 245-264.

Earman, J. (2008), `Reassessing the prospects for a growing block model of the universe', %%@
{\em International Studies in the Philosophy of Science} {\bf 22}, pp. 135-164.

Earman, J. (2008a), `Pruning some branches from ``branching spacetimes'' ', in D. Dieks %%@
(ed.) {\em The Ontology of Spacetime II}, Amsterdam: Elsevier, pp. 187-205.

Erhart, J., Sponar S., Sulyok G. et al. (2012), `Experimental demonstration of a %%@
universally valid error-disturbance uncertainty relation in spin measurements', {\em Nature %%@
Physics}, 15 January 2012,  pp. 1-5, DOI: 10.1038/NPHYS2194.

Galapon, E. (2002), `Self-adjoint time operator is the rule for discrete semi-bounded %%@
Hamiltonians', {\em Proceedings of the Royal Society of London} {\bf A 458}, pp. 2671-2689.

Greaves, H. (2010), `Towards a geometrical understanding of the CPT theorem', {\em British %%@
Journal for the Philosophy of Science}, {\bf 61}, pp. 27-50.

Greaves, H. and Thomas T. (2012), `The CPT theorem', arxiv:1204.4674.

Halliwell, J. (2011), `Decoherent histories analysis of minisuperspace quantum cosmology', %%@
{\em Journal of Physics Conference Series} {\bf 306}, 012023; (Proceedings of the 2010 DICE %%@
Conference); arxiv: 1108.5991

Hegerfeldt, G. and Muga, J. (2010), `Symmetries and time operators', {\em Journal of %%@
Physics A: Mathematical and Theoretical} {\bf 43}, p. 505303 (18 pages).

Hilgevoord, J. (1996), `The  uncertainty principle for energy and time I', {\em American %%@
Journal of Physics} {\bf 64}, pp. 1451-1456.

Hilgevoord, J. (1998), `The  uncertainty principle for energy and time II', {\em American %%@
Journal of Physics} {\bf 66}, pp. 396-402.

Hilgevoord, J. (2002), `Time in quantum mechanics', {\em American Journal of Physics} {\bf %%@
70}, pp. 301-306.

 Hilgevoord, J. (2005), `Time in quantum mechanics: a story of confusion', {\em Studies in %%@
the History and Philosophy of Modern Physics} {\bf 36}, pp. 29-60.

Hilgevoord, J. and Uffink J. (1988), `The mathematical expression of the uncertainty %%@
principle', in  A. van der Merwe, F. Selleri and G. Tarozzi (ed.s), {\em Microphysical %%@
Reality and Quantum Formalism} (Proceedings of a Conference at Urbino, 1985), Dordrecht: %%@
Kluwer, pp. 91-114.

Hilgevoord, J. and Uffink J. (1990), `A new view on the uncertainty principle', in  A.I. %%@
Miller (ed.), {\em Sixty-two Years of Uncertainty: Historical and Physical Inquiries into %%@
the Foundations of Quantum Mechanics}, New York: Plenum, pp. 121-139.

Hilgevoord, J. and Uffink J. (2006), `The  uncertainty principle', {\em Stanford %%@
Encyclopedia of Philosophy}: http://www.seop.leeds.ac.uk/entries/qt-uncertainty/

Hilgevoord, J. and Atkinson, D. (2011), `Time in quantum mechanics', in Callender (2011), %%@
pp. 647-662.

Holland, P. (1993), {\em The Quantum Theory of Motion}, Cambridge University Press.

Huggett, N.  and Wuthrich C. (this volume), `Time in Quantum Gravity'.

Isham, C. (1995), {\em Lectures on Quantum Theory: mathematical and structural %%@
foundations}, Imperial College Press.

Jammer, M. (1974), {\em The Philosophy of Quantum Mechanics}, John Wiley.

Jauch, J. (1968), {\em The Foundations of Quantum Mechanics}, Addison-Wesley.

Johns, O. (2005), {\em Analytical Mechanics for Special Relativity and Quantum Mechanics}, %%@
Oxford: University Press.

Kiefer, C. (2011), `Time in quantum gravity', in Callender (2011), pp. 663-678.

Kiukas, J., Ruschhaupt A., Schmidt P. and R. Werner (2011), `Exact energy-time uncertainty relation for arrival time by absorption', arxiv: 1109.5087v1

Landau, H. and Pollak H. (1961), `Prolate spheroidal wave functions: Fourier analysis and %%@
uncertainty II', {\em Bell Systems Technical Journal}, {\bf 40}, pp. 63-84.

Lewis, D. (1986), {\em On the Plurality of Worlds}, Oxford; Blackwell.

Maassen, H. and Uffink J. (1988),  `Generalized entropic uncertainty relations', {\em %%@
Physical Review Letters} {\bf 60}, pp. 1103-1106.

Malament, D. (2004), `On the time reversal invariance of classical electromagnetic theory', %%@
{\em Studies in the History and Philosophy of Modern Physics}, {\bf 35}, pp. 295-315.

McCall, S. (2000), `QM and STR: the combining of quantum mechanics and relativity theory', %%@
{\em Philosophy of Science}, {\bf 67}, pp. S535-S548.

Muga, J., Sala Mayato, R. and Egusquiza I. (eds.) (2008), {\em Time in Quantum Mechanics}, %%@
Berlin; Springer, two volumes.

North, J. (2008), `Two notions of time reversal', {\em Philosophy of Science}, {\bf 75}, %%@
pp. 201-223.

Ozawa, M. (2003), `Universally valid reformulation of the Heisenberg  uncertainty principle %%@
on noise and disturbance in measurements', {\em Physical Review A} {\bf 67}, 142105.

Ozawa, M. (2004), `Uncertainty relations for noise and disturbance in generalized quantum %%@
measurements', {\em Annals of Physics} {\bf 311}, pp. 350-416.

Placek, T. (2000)), `Stochastic outcomes in branching spacetime: analysis of Bell's %%@
theorems', {\em British Journal for the Philosophy of Science} {\bf 51}, pp. 445-475.

Reisenberger, M. and Rovelli C. (2002), `Spacetime states and covariant quantum theory', {\em Physical Review D}, {\bf 65}, 125016.

Rovelli, C. (2006), `Quantum gravity', in Part B of: J. Butterfield and J. Earman (eds.), %%@
{\em Philosophy of Physics}, in the series {\em The Handbook of Philosophy of Science}, %%@
Amsterdam: Elsevier, pp. 1287-1330.

Rovelli, C (2009), `Forget time', Essay for FQXi contest on the Nature of Time; arxiv: 0903.3832v3 

Saunders S.,  Barrett J.,  Kent A. and  Wallace D. (eds.) (2010), {\em Many Worlds? %%@
Everett, Quantum Theory and Reality}, Oxford: Oxford University Press.

Struyve, W.  (2010),   `Pilot-wave theory and quantum fields', {\em Reports on Progress in %%@
Physics} {\bf 73} , pp. 106001; arxiv: 0707.3685v4.

Struyve, W.  (2011),   `Pilot-wave approaches to quantum field theory', {\em Journal of %%@
Physics: conference series} {\bf 306}, pp. 012047; arxiv: 1101.5819.

Tooley, M. (1997), {\em Time, Tense and Causation}, Oxford: Oxford University Press.

Uffink, J. (1990), {\em Measures of Uncertainty and the Uncertainty Principle}, Utrecht %%@
University PhD. Available at: http://www.projects.science.uu.nl/igg/jos/

Uffink, J. and Hilgevoord, J. (1985), `Uncertainty principle and uncertainty relations', %%@
{\em Foundations of Physics} {\bf 15}, pp. 925-944.

Uffink, J. (1993), `The rate of evolution of a quantum state', {\em American Journal of %%@
Physics} {\bf 61}, pp. 935-936.

Wallace, D. (2001), `Worlds in the Everett interpretation', {\em Studies in the History and %%@
Philosophy of Modern Physics} {\bf 33}, pp. 637-661.

Wallace D. (2012), {\em The Emergent Multiverse}, Oxford: Oxford University Press.

Wallace, D. (2012a), `Decoherence and its role in the modern measurement problem', %%@
forthcoming in a special issue of {\em Philosophical Transactions of the Royal Society}; %%@
http://uk.arxiv.org/abs/1111.2187

Wallace, D. (this volume), `Arrows of time'.

Wuthrich, C. (this volume), `Time in quantum gravity'.

Yearsley, J. (2011), `Aspects of time in quantum theory', PhD thesis, Imperial College %%@
London; available at: arxiv.org: 1110.5790

\end{document}